\documentclass[aps,prl,twocolumn,showpacs,superscriptaddress,groupedaddress]{revtex4}  
\usepackage{graphicx}  
\usepackage{dcolumn}   
\usepackage{bm}        
\usepackage{amssymb}   

\begin{document}

\widetext


\title{Desingularization of the Milne Universe}

\author{Chethan Krishnan}\email{chethan@cts.iisc.ernet.in}
\author{Shubho Roy}\email{sroy@cts.iisc.ernet.in}
\affiliation{Center for High Energy Physics, Indian Institute of Science, Bangalore, India}

\date{\today}

\begin{abstract}
Resolution of cosmological singularities is an important problem in any full theory of quantum gravity. The Milne orbifold is a cosmology with a big-bang/big-crunch singularity, but being a quotient of flat space it holds potential for resolution in string theory. It is known however, that some perturbative string amplitudes diverge in the Milne geometry. 
Here we show that flat space higher spin theories can effect a simple resolution of the Milne singularity when one embeds the latter in 2+1 dimensions. We explain how to reconcile this with the expectation that non-perturbative string effects are required for resolving Milne.
Along the way, we introduce a Grassmann realization of the \.{I}n\"on\"u-Wigner contraction to export much of the AdS technology to our flat space computation.
\end{abstract}

\maketitle

\section{Introduction and Conclusion}

General Relativity is expected to require modifications at short distances. The oft-stated reason for this expectation is the existence of an infinite number of perturbative UV divergent couplings when one quantizes metric fluctuations. String theory solves this problem because it has an enormous gauge symmetry, called worldsheet conformal invariance. This gauge symmetry of string theory, essentially uniquely fixes the  infinite number of couplings arising in perturbative gravity. 

Apart from the quantum problem of divergences, there is also a purely classical reason why we expect that gravity might require modifications at short distances. This is because in gravity, spacetime singularities are ubiquitous \cite{PenHawk}. Since string theory is expected to be perturbatively finite in the UV, it is natural to wonder whether it can also resolve spacetime singularities. Some progress along this direction, and answers in the affirmative of various degrees of strength, can be found in \cite{Strominger, McGreevy, UCSB}.

Singularities in cosmological (a.k.a. time-dependent) spacetimes are especially tricky in string theory because typically we only understand how to quantize string theory in supersymmetric backgrounds, and supersymmetric backgrounds are automatically time {\em in}dependent. One way forward is to consider cosmological quotients of flat space as simple examples of time dependent singular backgrounds. The idea is that since the covering space is flat, we should be able to use some of the standard tools from flat space string theory, to explore these singular geometries. A simple context where a lot of papers on this topic \footnote{There is a whole slew of papers written on this topic in the context of singularity resolution in string theory, so our citation list is necessarily incomplete.} have been written is the case of the Milne orbifold, which is a time dependent orbifold of flat space (See \cite{Cornalba1, Cornalba2, Cornalba3, Seiberg1, Seiberg2, Nekrasov, Craps1, Craps2} for related work.). It turns out that some tree level string scattering amplitudes are singular on the Milne orbifold \cite{Pioline1, Pioline2, CrapsReview}, indicating that perturbative string theory breaks down. Also, because it is an exact CFT, there are no $\alpha'$-effects that can result in a resolution of the Milne singularity \footnote{However, for a proposal that winding tachyon condensation can resolve singularities, see eg. \cite{McGreevy}.}. Together, if we take these two statements at face value, we come to the conclusion that only non-perturbative $g_s$-effects can come to the rescue of Milne, perhaps in a context like the AdS/CFT duality \footnote{We thank Ben Craps and Boris Pioline for correspondence on related questions.}.



In this paper, we will study the Milne orbifold from another perspective. We will consider it in the context of Chern-Simons higher spin theories in 2+1 dimensional flat space \cite{Bagchi, Troncoso}. Our motivation is as follows. 
It is expected \cite{Sundborg, Witten} that higher spin theories capture features of string theory in the tensionless limit \footnote{See \cite{Arjun2} for a connection between tensionless strings and flat space.}, which corresponds to $\alpha' \rightarrow \infty$. Therefore heuristically, the gauge symmetries of (classical) higher spin theory can be thought of as a target space realization of the worldsheet gauge symmetries of tree-level string theory in the $\alpha' \rightarrow \infty$ limit. So in this limit it is possible to ask whether spacetime singularities are artifacts of a singular gauge, and if so, whether one can get rid of them by going to a different gauge. We will indeed see that by doing a flat space higher spin gauge transformation, we are able to remove the singularity in the Milne Universe in a very simple and natural way. This is our main result. To accomplish this, we use a Grassmann trick to rewrite the flat space Chern-Simons HS theory in a form that closely resembles the AdS case. This trick in itself is pretty powerful, but seems to have gone unnoticed before.

Note that in this picture the string coupling $g_s \sim 1/N$ never showed up. There is superficially some tension between this and the general belief from string theory that non-perturbative effects are necessary for resolving Milne 
\footnote{The connection between string theory and higher spin theory (in particular C-S higher spin theory in flat 2+1 D space) has not been made precise, and is mostly heuristic at present. So it is not entirely clear that our result is a bona-fide challenge to the string theory expectations.}. 

Perhaps the best way to understand our result is to note that the limit $\alpha' \rightarrow \infty$ is precisely the opposite of the limit where the usual Einstein gravity emerges in string theory 
($\alpha' \rightarrow 0$). That is, higher spin theories are a different classical limit of string theory. In this limit, the stringy gauge invariances have a simple target space realization in terms of higher spin gauge symmetries. So what we do here amounts more precisely to a de-singularization via a gauge transformation, and not to a resolution \footnote{Loosely, we will use the two terms interchangeably.}: the latter is usually accomplished via the addition of new degrees of freedom, and that is the situation that is envisaged in the usual discussions of the Milne orbifold. We emphasize however that it is not that the gauge transformation here is singular, it is that the solution (in the metric language) has interpretation as a spacetime singularity \cite{Craps2}. The Chern-Simons gauge field is in fact regular before and after the resolution, even though the metric is non-singular only after \footnote{It is perhaps also worth emphasizing that the Chern-Simons formulation of gravity and higher spin gravity that we are using is best suited for classical questions. Note that in the spin-2 theory (the usual C-S gravity) it is currently believed that the quantization of the gravity is {\em not} the same as the quantization of the C-S theory \cite{Wittenfoot}. It is best thought of as only a classical equivalence. This is because invertibility of the vielbeins is required for the CS interpretation, so in the path integral one is integrating over different field configurations. Another relevant observation is that the existence of black holes (at least in the AdS$_3$ context), indicates a huge degeneracy of states which is surprising in a Chern-Simons theory. We will be using the holonomies to distinguish classical solutions, not define observables in the quantum theory.}.

Some recent papers dealing with cosmologies and singularities in a higher spin set up can be found in \cite{CK1, CK2, Pando}. In \cite{CK2} a cosmological sinularity resolution was done, but in the context of dS$_3$ (higher spin) gravity. The reason why Milne is of much more interest than the dS quotients that we considered in the previous paper is because the geometry is locally flat here, so one can potentially consider string propagation on it. Indeed, Milne has been studied rather extensively in a stringy context as already mentioned. By contrast the singularity we resolved in dS, is an obscure and essentially unknown one, and was merely interesting as a proof of principle. We could not resolve Milne at the time, because flat space higher spin theories were constructed only afterwards \cite{Bagchi, Troncoso}. 

Indeed, after the first version of this paper appeared, we have revisited the 2-to-2 string scattering amplitude on the Milne obifold, and exhaustively scanned for divergences \cite{Ayush}. The result is that {\em all} the singularity-related divergences arise when the $\alpha'$ (made dimensionless by multiplication with appropriate momenta) is less than some numerical value. The remaining divergences all arise when the tower of intermediate string states goes on-shell, and are physical IR divergences. Since higher spin theories are morally the $\alpha' \rightarrow \infty$ limit of string theory, we believe the fact that we are able to resolve the geometry at infinite $\alpha'$ and the fact that string amplitudes are UV finite when $\alpha'$ is large enough, is at the very least, suggestive.

\section{Flat 2+1 D (Higher Spin) Gravity} 

We will work with the spin-3 theory in this paper. In 2+1 dimensions, working with higher spin theories is easily acoomplished via the Chern-Simons formulation of gravity \cite{WittenCS}, but with a higher rank version of $ISO(2,1)$ as discussed in \cite{Bagchi, Troncoso}. 

A lot of work on higher spin theories has been done in the context of AdS$_3$ theories, and we will make an observation that enables us to translate a lot of this AdS machinery to the flat space theory. This observation is that if one makes the replacement 
\begin{equation}
\frac{1}{\ell} \rightarrow \epsilon
\end{equation}
where $\ell$ is the AdS$_3$ radius and $\epsilon$ is a Grassmann parameter defined by the condition that $\epsilon^2=0$, then the AdS Chern-Simons action \footnote{See for example Section 2 and appendix A of \cite{Maloney}} written in terms of the triad and the spin connection (and their higher spin cousins) reduces to the flat space Einstein-Hilbert action (and its higher spin cousin) times 
$\epsilon$, provided one takes the Newton's constant to be
\begin{eqnarray}
\frac{1}{16G}=k. 
\end{eqnarray}
Since we are only concerned about classical equations and their solutions, the overall Grassman factor in the action will not affect our discussions.
The basic reason why this trick works is because of the fact that $ISO(2,1)$ is the \.{I}n\"on\"u-Wigner contraction of $SL(2,R) \times SL(2,R)$ (and similarly, for the higher spin generalizations). This approach and some of its applications are further explored in \cite{KRNEW}. 

There are two basic reasons why this trick is useful. 
\begin{itemize}
\item We can adopt the notations and the $SL(3)$ generator matrices of \cite{Maloney} without modifications as long as we make sure that $1/\ell \rightarrow \epsilon$ squares to zero. Without the Grassmann approach, we would be faced with the task of constructing an explicit set of matrices for the generators in \cite{Bagchi, Troncoso}, such that they have a non-degenrate trace form.
\item The non-degenerate trace form of the AdS Chern-Simons theory descends to give us a non-degenerate trace form for the flat space theory with this trick. 
\end{itemize}

The upshot is that we can work with flat space (higher spin) gravity in 2+1 dimensions by starting with the AdS Chern-Simons theory, writing the gauge field in terms of the vielbein and spin connection, and seting $1/\ell \rightarrow \epsilon$ with $\epsilon^2=0$. 


\section{Milne: Metric and Connection} 

We will take the Milne metric in 2+1 dimensions in the form \cite{Barnich:2012aw,Barnich:2012xq}
\begin{equation}
ds^{2}=-dT^{2}+r_{C}^{2}dX^{2}+\alpha^{2}T^{2}d\varphi^{2},\label{eq: Milne metric}
\end{equation}
where for comparison with \cite{BagchiOld} we define the metric parameters $\alpha,r_{C}$ in terms of the ``mass'',
$M$ and ``spin'', $J$ by, 
\begin{eqnarray}
\alpha=\sqrt{M},\qquad r_{C}=\sqrt{\frac{J^{2}}{4M}.}
\end{eqnarray}
\label{The-Milne-metric}(We are following the convention, where $8G=1$).
$X,\varphi$ directions are compact and closed, both with period $2\pi$. The spacetime looks like a double cone. There is a causal structure singularity at $T=0$ which is where the $\varphi$-circle crunches before re-expanding in a big-bang.
The geometry can be understood in terms of an orbifold of flat Minkowski space and that is what makes it tractable in string theory, but we will not elaborate it here. We have taken the $X$-direction to be compact to match with some of the flat space holography literature, our results are essentially unchanged even if we drop this assumption.

The corresponding expressions for the triads and spin-connection (dualized using $\epsilon_{abc}$)
are,
\begin{eqnarray}
e^{0}  =  dT, \ \ 
e^{1}  =  r_{C}dX, \ \ 
e^{2}  =  \alpha Td\varphi,\label{eq:Milne triads}
\end{eqnarray}
\begin{eqnarray}
\omega^{0}  =  0,\ \ 
\omega^{1}  =  \alpha d\varphi,\ \ 
\omega^{2}  =  0.\label{eq: Milne dual spin connection}
\end{eqnarray}
So in the Chern-Simons language the $SL(2)$ Grassmann valued connection
is \footnote{Note that the $1/\ell$ in the AdS case \cite{Maloney} has been repleaced with the Grassmann paramter $\epsilon$.},
\begin{eqnarray}
A^{\pm}&=&\left(\omega^{a}\pm\epsilon\: e^{a}\right)T_{a} \label{eq: Milne connection}\\
&=&\pm\left(\epsilon\: dT\right)T_{0}+\left(\alpha d\varphi\pm\epsilon\: r_{C}dX\right)T_{1}\pm\left(\epsilon\:\alpha Td\varphi\right)T_{2}. \nonumber
\end{eqnarray}
The $\varphi$-circle holonomy is,
\begin{eqnarray*}
W_{\varphi}^{\pm} & \equiv & P\exp\left(\oint\: d\varphi\: A_{\varphi}^{\pm}\right)\\
 & = & \exp\left[2\pi\alpha\left(T_{1}\pm\epsilon\: T\: T_{2}\right)\right].
\end{eqnarray*}
The eigenvalue spectrum of the holonomy matrix, $w_{\phi}^{\pm}\equiv2\pi\alpha\left(T_{1}\pm\epsilon T\: T_{2}\right)$
is $(0,\pm2\pi\alpha)$. Similarly we can compute the $X$-circle
holonomy, $W_{X}^{\pm}=P\exp\left(\oint\: dX\: A_{X}^{\pm}\right)=\exp w_{X}^{\pm}$,
$w_{X}^{\pm}=\pm\epsilon\;2\pi r_{C}T_{1}$. Its eigenvalues are $0,\pm2\pi r_{C}\epsilon.$
For future reference we also note the characteristic polynomial coefficients \cite{Maloney}
of these holonomy matrices (These are identical for both $\pm$ sectors,
so we drop the superscripts).
\begin{equation}
\Theta_{\varphi}^{0}\equiv\mbox{det}\left(w_{\varphi}\right)=0,\qquad\Theta_{X}^{0}\equiv\mbox{det}\left(w_{X}\right)=0,\label{eq:Theta zeroes for Milne}
\end{equation}
\begin{equation}
\Theta_{\varphi}^{1}=\mbox{tr}\left(w_{\varphi}^{2}\right)=8\pi^{2}\alpha^{2},\qquad\Theta_{X}^{1}=0.\label{eq: Theta ones for Milne}
\end{equation}

\section{Adding Higher Spins} 

To avoid clutter we work exclusively with the ``holomorphic" gauge field ($A^{+}$)
and drop the superscript. The ``antiholomorphic" ($A^{-}$) sector works
out exactly parallel. With that understood, to resolve/remove the singularity, now we turn on the higher spin sector:
\[
A'=A+\sum_{a=-2}^{2}\left(C^{a}+\epsilon\: D^{a}\right)W_{a}.
\]
We demand that $C^{a},D^{a}=C^{a}(T),D^{a}(T)$ for simplicity. If
we include $\phi,X$ dependence as well, then the path-ordered exponentials
for the holonomies become hard to evaluate. Besides, we hope to be able to resolve Milne without breaking the symmetries of the geometry. Fortunately, we are able to find a resolution while satisfying these restrictions.

The metric components resulting from inclusion of the higher spin
generators is then \cite{Campoleoni},
\begin{eqnarray}
g_{\mu\nu}' & = & g_{\mu\nu}+\frac{1}{2}D_{(\mu}^{a}D_{\nu)}^{b}Tr\left(W_{a}W_{b}\right)\nonumber \\
 & = & g_{\mu\nu}+\frac{4}{3}D_{\mu}^{0}D_{\nu}^{0}-2D_{\mu}^{1}D_{\nu}^{-1}
 -2D_{\mu}^{-1}D_{\nu}^{1}+\hspace{0.5in}\nonumber\\
&& \hspace{1.3in}+8D_{\mu}^{2}D_{\nu}^{-2}+8D_{\mu}^{-2}D_{\nu}^{2}.\label{eq: spin 3 corrected metric components}
\end{eqnarray}
Here, $g_{\mu\nu}$ are the components of the Milne metric (\ref{eq: Milne metric}).

For this connection to describe the same gauge configuration, the
new holonomy matrices must be in the same conjugacy class. The new
holonomies are,
\[
w'_{\varphi}=2\pi\left(\alpha T_{1}\pm\epsilon\:\alpha T\: T_{2}+C_{\varphi}^{a}W_{a}+\epsilon D_{\varphi}^{a}W_{a}\right),
\]
\[
w'_{X}=2\pi\left(\epsilon\: r_{C}T_{1}+C_{X}^{a}W_{a}+\epsilon D_{X}^{a}W_{a}\right).
\]
The eigenvalues are hard to evaluate directly, so we work with the
characteristic polynomial coefficients instead,
i.e, $\Theta'\,_{\varphi,X}^{0}\equiv\mbox{det}\left(w'_{\varphi,X}\right),$
$\Theta'\,_{\varphi,X}^{1}\equiv\mbox{tr}\left(w'_{\varphi,X}\,^{2}\right)$.
These must be identical to (\ref{eq:Theta zeroes for Milne}-\ref{eq: Theta ones for Milne}),
for the new connection to describe the same gauge configuration.

For general $C$'s and $D$'s, these relations are complicated (but computable) algebraic expressions, so we will not present them in the general case. Our goal is merely to see whether we can come up with {\em some} resolution of the Milne singularity for some choice of $C$'s and $D$'s.

Of course, one needs the new connection to be flat so that it will be a solution to
the equations of motion.
\begin{equation}
F'=dA'+A'\wedge A'=0.
\end{equation}
Expanding $A'=A+C+\epsilon D$, and noting that $A$ is already flat, the flatness of $A'$ gives (in terms of  components)
\[
\left[C_{\mu},C_{\nu}\right]=0,
\]
\[
\partial_{\mu}C_{\nu}-\partial_{\nu}C_{\mu}+\left[C_{\mu},\omega_{\nu}\right]+\left[\omega_{\mu},C_{\nu}\right]=0,
\]
\[
\partial_{\mu}D_{\nu}-\partial_{\nu}D_{\mu}+\left[D_{\mu},\omega_{\nu}\right]+\left[\omega_{\mu},D_{\nu}\right]+\left[C_{\mu},e_{\nu}\right]+\left[e_{\mu},C{}_{\nu}\right]=0,
\]
\[
\left[C_{\mu},D_{\nu}\right]+\left[D_{\mu},C_{\nu}\right]=0.
\]

\section{The Milne Resolution}


We look at the simple ansatz, 
\begin{equation}
C^a_{\mu}=0. \label{ansatzC}
\end{equation}
This generates a holonomy condition from preserving $\Theta_{\varphi}^{0}$,
\begin{equation}
D_{\varphi}^{0}=3\left(D_{\varphi}^{2}+D_{\varphi}^{-2}\right),\label{eq: holonomy 1}
\end{equation}
while the rest of the holonomy conditions are automatically satisfied
by this ansatz.

Also, this ansatz automatically satisfies $3$ of the $4$ flatness
conditions and leads to the following equation of motion for $D$'s,

\[
\partial_{\mu}D_{\nu}-\partial_{\nu}D_{\mu}+\left[D_{\mu},\omega_{\nu}\right]+\left[\omega_{\mu},D_{\nu}\right]=0.
\]
This further leads to,
\begin{itemize}
\item $\partial_{T}D_{X}=0,$ i.e. $D_{X}^{a}$'s are constant.
\item $\left[D_{X},\omega_{\varphi}\right]=0,$ which sets, $D_{X}^{\pm1}=0,D_{X}^{\pm2}=-\frac{1}{2}D_{X}^{0}.$
\item $\partial_{T}D_{\varphi}+\left[D_{T},\omega_{\varphi}\right]=0$ .
\end{itemize}
This system of equations is simply solved by $D_{X},D_{T}=0$ and
$D_{\varphi}=\mbox{const.}$

In particular, $D_{\varphi}^{0}=3D_{\varphi}^{2}$ while $D_{\varphi}^{\pm1},D_{\varphi}^{-2}=0$
is a solution to the equations of motion with the same holonomy as
that of the Milne orbifold. The reason we picked the ansatz (\ref{ansatzC}) is that the resultant metric gets modified
{\em only} in its $\varphi\varphi$-component, 
\begin{equation}
g'_{\varphi\varphi}=g_{\varphi\varphi}+12\left(D_{\varphi}^{2}\right)^{2}.\label{eq: resoved g_phiphi}
\end{equation}
The curvature scalars are well-defined everywhere. We quote the resultant Ricci scalar,
\begin{equation}
\mathcal{R}=\frac{24\left(D_{\varphi}^{2}\right)^{2}\alpha^{2}}{\left(12\left(D_{\varphi}^{2}\right)^{2}+T^{2}\alpha^{2}\right)^{2}}\label{eq: Resolved Ricci}
\end{equation}
which is finite and continuous at $T=0$.

So the resolution that we have done here effectively ensures that the shrinking Milne circle has a minimum finite radius at the erstwhile singularity. Remarkably and satisfyingly, we are able to preserve all the symmetries of the original geometry in doing this resolution, and the metric now looks like a smooth bounce, instead of the Milne cone with a crunch/bang.


The original geometry did not have any higher spin fields, but after the gauge transformation, we need to also check that the resultant spin-3 field is regular as well. 
The spin-3 fields can be computed via  \cite{Campoleoni}
$\Phi_{\mu\nu\rho}=\frac{1}{9}\mbox{tr}\left(E_{(\mu}E_{\nu}E_{\rho)}\right)$.
For the resolved Milne orbifold the nonvanishing components are,
\begin{eqnarray*}
\Phi_{\varphi\varphi\varphi} & = & -\frac{16}{3}\left(D_{\varphi}^{2}\right)^{3}+\frac{4}{3}D_{\varphi}^{2}T^{2}\alpha^{2},\\
\Phi_{TT\varphi} & = & \frac{4}{9}D_{\varphi}^{2}, \\
\Phi_{TX\varphi} & = & -\frac{2}{9}D_{\varphi}^{2} r_C
\end{eqnarray*}
which is manifestly regular everywhere. We can think of these higher spin fields as the matter supporting the resolved Milne geometry.


\vspace{0.2in}

CK thanks Arjun Bagchi, Ben Craps, Justin David, Oleg Evnin, Jarah Evslin, Prem Kumar and Boris Pioline for discussions, and Chalmers Institute of Technology, Sweden and the Yerevan State Univeristy, Armenia (during the ``Quantum Aspects of Black Holes and its Recent Progress" worskhop) for hospitaility during parts of this work. The research of SR is supported by Department of Science and Technology (DST), Govt.$\hspace{0.05in}$of India research grant under scheme DSTO/1100 (ACAQFT).

\end{document}